# A high yield method for protoplast isolation and ease detection of rol B and C genes in the hairy roots of cauliflower (Brassica oleracea L.) inoculated with Agrobacterium rhizogenes

Qutaiba Shuaib Al-Nema
*Department of Biology, College of Education for Pure Sciences, University of Mosul*

Ghazwan Qasim Hasan
*Department of Biology, College of Education for Pure Sciences, University of Mosul*, dr.ghazwan@uomosul.edu.iq

Omar Abdulazeez Alhamd
*Department of Biology, College of Education for Pure Sciences, University of Mosul*

# A high yield method for protoplast isolation and ease detection of rol B and C genes in the hairy roots of cauliflower (Brassica oleracea L.) inoculated with Agrobacterium rhizogenes

## Abstract

Protoplasts represent a unique experimental system for the circulation and formation of genetically modified plants. Here, protoplasts were isolated from genetically modified hairy root tissues of Brassica oleracea L. induced by the Agrobacterium rhizogenes strain (ATCC13332). The concentration of enzyme solutions utilized for protoplast isolation was 1.5 % Cellulase YC and 0.1 % Pectolyase Y23 in 13% mannitol solution, which resulted in high efficiency of isolation within 8 hours, in which the protoplast yield was $2 \times 10^4$ cells ml-1 and the percentage of viability was 72%. Each protoplast has one nucleus with a nucleation of 48%. A polymerase chain reaction (PCR) assay verified the presence of rol B and rol C genes in hairy root tissues by detaching a single bundle of DNA replication from these roots using a specific pair of primers. The current study demonstrated that A. rhizogenes strain (ATCC13332) is a vector for the incorporation of T-DNA genes into cauliflower plants, as well as the success of the hairy roots retention of rol B and rol C genes transferred to it.

## Keywords

## Creative Commons License

RESEARCH PAPER

# A High Yield Method for Protoplast Isolation and Ease Detection of *rolB* and *C* Genes in the Hairy Roots of Cauliflower (*Brassica oleracea* L.) Inoculated with *Agrobacterium rhizogenes*

Qutaiba Shuaib Al-Nema, Ghazwan Qasim Hasan*, Omar Abdulazeez Alhamd

Department of Biology, College of Education for Pure Sciences, University of Mosul, Iraq

## Abstract

Protoplasts represent a unique experimental system for the circulation and formation of genetically modified plants. Here, protoplasts were isolated from genetically modified hairy root tissues of *Brassica oleracea* L. induced by the *Agrobacterium rhizogenes* strain (ATCC13332). The concentration of enzyme solutions utilized for protoplast isolation was 1.5% Cellulase YC and 0.1% Pectolyase Y23 in 13% mannitol solution, which resulted in high efficiency of isolation within 8 hours, in which the protoplast yield was $2 \times 10^4$ cells ml$^{-1}$ and the percentage of viability was 72%. Each protoplast has one nucleus with a nucleation of 48%. A polymerase chain reaction (PCR) assay verified the presence of *rol B* and *rolC* genes in hairy root tissues by detaching a single bundle of DNA replication from these roots using a specific pair of primers. The current study demonstrated that *A. rhizogenes strain* (ATCC13332) is a vector for the incorporation of T-DNA genes into cauliflower plants, as well as the success of the hairy roots retention *of rol B* and *rolC* genes transferred to it.

## 1. Introduction

Cauliflower, also locally known as Al-Zahra or Shaflur, is an essential edible vegetable that belongs to *Brassica oleracea*, a species of the genus *Brassica* within the Brassicaceae family [1,2]. Only the white head is eaten in most cases, and it is sometimes referred to as "curd" [3]. *B.oleracea* also involves broccoli, cabbage, kale, Brussels sprouts, Kohlrabi, and many others, all called "cole" crops [4]. Transgenic hairy root cultures arising from *Agrobacterium rhizogenes* inoculation are among the most important methods for producing transgenic plants. Hairy root cultures (HRCs) inoculated with *A. rhizogenes* have been used to treat a wide range of plants [5]. For example, Transgenic *Solanumnigrum* L. plants could emerge spontaneously from these induced roots [6]. Moreover, the roots induced by *A. rhizogenes* 1601 harbouring Ri-plasmid also stimulated the formation of transgenic callus from sugar beet plants [7]. In addition, the protoplasts were isolated directly from *Maesalanceolata* L. roots induced by *A. rhizogenes* strain (LBA 9402/12) showed continued to divide and transform into a callus [7]. Furthermore, rather than including the genetic material directly into the protoplast cells, the genetic transformation can be accomplished by fusing the transgenic hairy root protoplast with the receiving protoplast and producing transgenic plants [8]. The goal of this study is to obtain a transgenic protoplast that can be utilized for somatic hybridization and the production of genetically modified plants, as well as to confirm the presence *of rol B* and *rol C* genes in transgenic hairy root tissues by PCR technique.

## 2. Materials and methods

### 2.1. Bacterial strain

The bacterial strain ATCC13332 was prepared from non-transgenic *A. rhizogenes* containing the pRi plasmid from the Leibniz Institute DSMZ (German Collection of Microorganisms and Cell Cultures GmbH). The bacteria were cultivated overnight in a Nutrient broth medium at 28 °C.

### 2.2. Plant materials

*B.oleracea* seeds were sterilized by immersing them in 70% ethyl alcohol for 1 min. Next, the seeds were immersed in sodium hypochlorite solution (commercial minor at a concentration of 6%) at a ratio of 1 volume of sodium hypochlorite (NaOCl) to 1 volume of sterile water for 10 min [9]. The seeds were sowed on MSO medium after being washed three times in sterile water and stored in the culture room under alternating light conditions of 16 hours of light and 8 hours of darkness (2000 lux, 25 ± 2 °C) [10].

### 2.3. Agrobacterium inoculation

At the age of 15 days, stems were cut to 2–3 cm long and leaves were removed from seedlings. Stems were directly injected at four different sites using a micro-needle (ICC Insulin Syringe U-100-29 G1/2–0.3314M) with 10 μl of an overnight culture of bacteria [11]. Leaves were also inoculated by injecting them from the bottom surface at two or more sites in the mid vein (central vein of a leaf). Explants were vertically infused into agar-solidified MSO and WP medium (2 pieces/jar). Samples were stored in the culture room at 25°C in the dark.

### 2.4. Protoplast isolation

As listed (Table 1), five enzymatic solutions for protoplast isolation [8,12] were prepared by dissolving the required concentrations of their powders in distilled water in the presence of 5% mannitol (pH 5.8).

One gram of healthy and young hairy roots was placed in 5.0 ml of CPW salt solution [13] containing 13% mannitol in a 9 cm diameter Petridish and incubated for 1 hour in dark conditions at 25 ± 2 °C. Following, 10 ml of the enzyme mixture was added to the incubated samples, and the dishes were kept at 25 ± 2 °C, 100 Lux [12]. Protoplast release was observed under a light microscope following 8 hours of incubation. The enzyme solution containing protoplast was passed through a 45 μm nylon sieve, then centrifuged at 100 g for 5 min, and washed twice in 5.0 ml of CPW13M solution by resuspension and centrifugation [13].

### 2.5. Protoplast yield and viability

0.1 ml of isolated protoplast suspension was placed on the Fuchs Rosenthal haemocytometer chamber, then the number of cells in 1.0 ml was counted and the total number of cells was estimated at the magnification of 10X and 40X [14]. Protoplast viability was estimated by Evans blue day solution, where 1.0 ml of protoplast suspension was mixed with 1.0 ml of 0.5% of day solution. The mixture was left at room temperature for 10 min, the cells were counted for viability using a haemacytometer chamber [15].

### 2.6. Staining of the protoplast nuclear

To visualize the protoplast nuclear, fifty μl of carbol fuchsin dye solution was mixed with twenty μl of isolated protoplast [16] and examined under a light microscope.

### 2.7. Confirmation of genetic transformation

Genomic DNA was extracted from the genetically modified hairy roots of *Brassica oleacea* according to the extraction protocol suggested by Healey et al. [17]. The purity and concentration of DNA were determined spectrophotometrically at both wavelengths of 260–280 nm [18]. A PCR reaction was performed to confirm the presence of the *rol B* and *rol B* genes in transformed hairy roots. Different specific pair primers with their were used to amplify the above genes [19]. Sequences of oligonucleotide primers used, PCR mixture and PCR conditions are listed in Table 2, Table 3 and Table 4, respectively.

### 2.8. Statistical analysis

The experiment was statistically analyzed, and the comparison of the significant differences between sample rates was accomplished using Duncan's new multiple range test (DMRT) [20].

Table 1. Enzyme solutions used for protoplasts isolation from hairy roots of Brassica oleracea.

| Enzymes | Mixtures | | | | |
| --- | --- | --- | --- | --- | --- |
| | I | II | III | IV | V |
| Cellulase Yc | 0 | 1.5 | 0.5 | 1.0 | 0 |
| Cellulase R10 | 0.5 | 0 | 0.5 | 0 | 1.0 |
| Cellulase RS | 0.5 | 0 | 0 | 0.5 | 0 |
| Pectolyase Y-23 | 0.1 | 0.1 | 0.1 | 0 | 0.1 |

Table 2. Sequences of oligonucleotide primers used in PCR reaction (5′ to 3′ orientation).

| Gene name | Primer | Sequence |
|---|---|---|
| rol B | F | 5′ATGGATCCCAAATTGTATTCCTTCCACGA 3′ |
| rol B | R | 5′TTAGGCTTCTTTCTTCAGGTTTACTGCAGC 3′ |
| rol C | F | 5′ CATTAGCCGATTGCAAACTTG 3′ |
| rol C | R | 5′ ATGGCTGAAGACCTG 3′ |

## 3. Results

### 3.1. Hairy root formation

The preliminary results showed the success of the inoculation of the leaves and stems of *B. oleracea* by *A. rhizogenes* ATCC13332 that were injected at more than one site. The induction of hairy roots began in the form of transverse roots from the inoculation areas of stems (Fig. 1 A), and their emergence in the mid vein from the lower surface of leaves (Fig. 1 B). Hairy roots are formed in the form of short, filamentous structures that are dense with root hairs (Fig. 1C).

In general, hairy roots were formed first in the inoculation areas, and the response of the leaves was earlier and better than the rest of the stem, and their emergence was in the form of short, minute, white roots in the inoculation areas, with two to four roots for each inoculation region. The emergence of hairy roots from areas that have not been inoculated and for a period similar to the period of their emergence from inoculated areas later developed into a dense group of roots with abundant white hairs. The findings showed that good cultures of hairy roots were obtained by growing them on WP medium and then ridding them of the bacteria by transferring them in successive transfers onto solid WP medium with graded concentrations of cefotaxime and remaining for 15 days (Fig. 1 D). The recovery of these cultures from the bacteria took more than two months.

### 3.2. Protoplast isolation

The results revealed the success of isolating protoplasts from hairy roots and the efficiency of the enzyme solution used (1.0% cellulase YC and 0.1% PectolyaseY-23 in the presence of 13% mannitol). The isolation process spent 8 hours with this solution, and swelling of the ends of root hairs was observed during the first hour (Fig. 2 A), followed by the release of a few protoplasts from a single root (Fig. 2 B and C). The absence of chloroplasts was observed in this type of protoplast (Fig. 2 D). The results also showed a significant increase in sample number 2 compared with others in the yield of protoplasts, viability, and nucleation of genetically transformed hairy roots, except the percent of nucleation, which showed no significant difference between samples 2 and 3 (Table 5). In addition, the transgenic protoplast had its vitality appear in a dormant color when stained with Evan's blue dye and viewed microscopically under a light microscope.

In contrast, the non-vital protoplast appeared in blue. Moreover, such protoplasts have nucleated, which is evident in their red color. Each protoplast contains one nucleus.

### 3.3. Hairy roots tissue retains rol B and rol C genes

The result of the gel electrophoresis assay of PCR samples for the transgenic hairy roots showed the detachment of a single bundle resulting from the replication DNA. These bundles with the expected size confirm the presence of *rol B* and *rol C* genes in the genome of transgenic hairy roots and the absence of separation of such bundles from the replicated DNA isolated from normal roots (Fig. 3).

## 4. Discussion

Our results showed that *A. rhizogenes* strain (ATCC13332) is a suitable vector for the incorporation of T-DNA genes into *B. oleracea* plants, and the success of transferring the *rol B* and *rol C* genes to the hairy roots. The utilization of the strain *A. rhizogenes* ATCC13332 to accomplish genetic transformation in Cauliflower plants explains that the wild type of strain did not make any modifications to its genetic repertoire, including genetic markers. In other words, it is considered a natural engineer [21]. The virulence factor contained in this wild strain of *A. rhizogenes*, the genome of the host plant, and the induction of this bacterium with chemical attraction compounds related to plant cell wall components to the initial bacterial-cell association before T-DNA transfer are all factors in the success of genetic transformation by this bacterium [22]. Hairy root induction is influenced by several factors such as strain type, bark density, plant type, and medium composition [23,24]. Studies of isolating protoplasts from the transformed hairy roots are still in their narrowest limits due to the possibility of contamination with the bacterial vector. In this study, the success of isolating protoplasts from genetically transformed hairy root tissues is due to the role of CPW 13M solution in the plasmolysis of root tip cells. The enzymatic mixture consisting of 1.0% cellulose YC and 0.1% pectolyase Y23 solution was suitable for digesting the tops of root hairs, removing their cell wall, and releasing protoplast with suitable levels of viability and density that could be grown [12]. In fact, the appropriateness of the osmotic pressure in the isolation conditions of this protoplast explains why it preserved its viability and spherical shape when its plasma membrane was exposed to ambient conditions and did not explode, despite the peculiarity of the tissues from which it was isolated [25]. It seems that appropriate expression is necessary for the growth of hairy roots, especially since the results of the PCR confirmed that the level of expression in *rol B* genes is high and that the level of high or low expression is associated with weak or active hairy root growth [26]. The level of expression shown by the *rol C* gene may have a stimulating effect on the production of secondary metabolites and encoding proteins of *rol A*, which is a transcription factor for gibberellin metabolism [27].

Table 3. PCR master mix for 24 μl reactions.

| Compounds | Volume μl | Final concentrations |
|---|---|---|
| Premix (Mgcl2 buffer 10x + Kcl + Tris − Hcl + dNTPs + Taq DNA Polymerase | 9.0 | ……… |
| Forward Specific Primer | 1.0 | 10 pmol |
| Reverse Specific Primer | 1.0 | 10 pmol |
| Template DNA | 4.0 | 50 ng |
| Double distilled water (ddH2O) | 9.0 | ……… |
| Final volume | 24 | ……… |

Table 4. PCR condition.

| Step | Temp. (°C) | Time (min.) | Cycles |
|---|---|---|---|
| Initial denaturation | 95 | 10 | 1 |
| Denaturation | 95 | 1 | |
| Annealing | rol B 62 | 1 | 35 cycles for each gene |
| | rol C 57 | | |
| Extension | 72 | 1 | |
| Final extension | 72 | 5 | 1 |

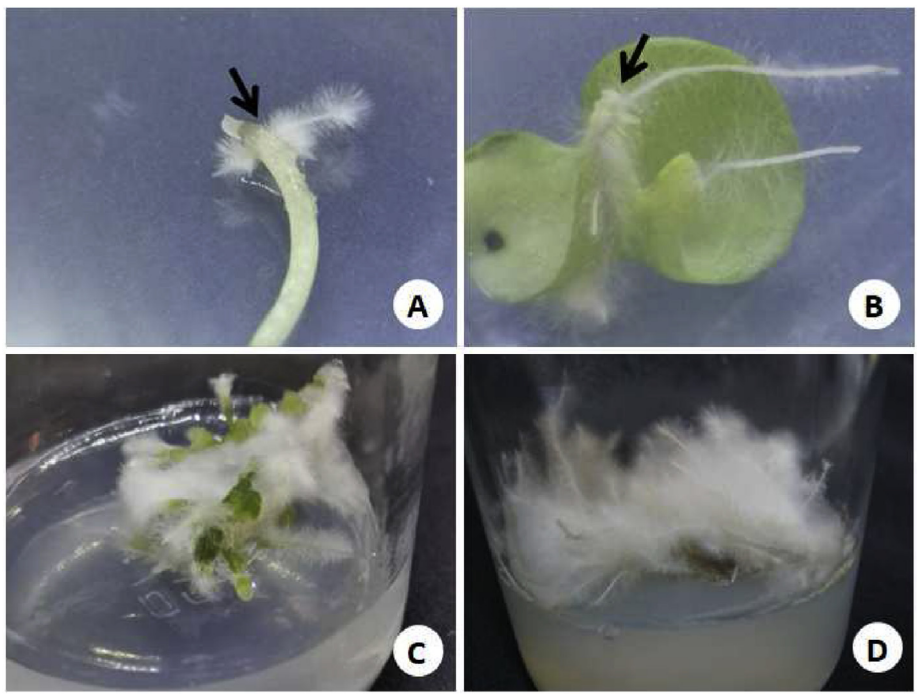

Fig. 1. Induction of hairy roots on explants of Brassica oleracea inoculated by direct injection with Agrobacterium rhizogenes ATCC 13332. A. Induced hairy roots (Black arrow) on the stem at the age of 9 days of growing on MSO medium; B. Induced hairy roots (Black arrow) on the stem at the age of 7 days of growing on MSO medium; C. Hairy roots are induced from the mid vein of a leaf at 15 day-olds growing in a solid MSO medium; D. Culture of 25-day-old cured hairy roots (C) growing in solid WP medium (note their density and negative growth).

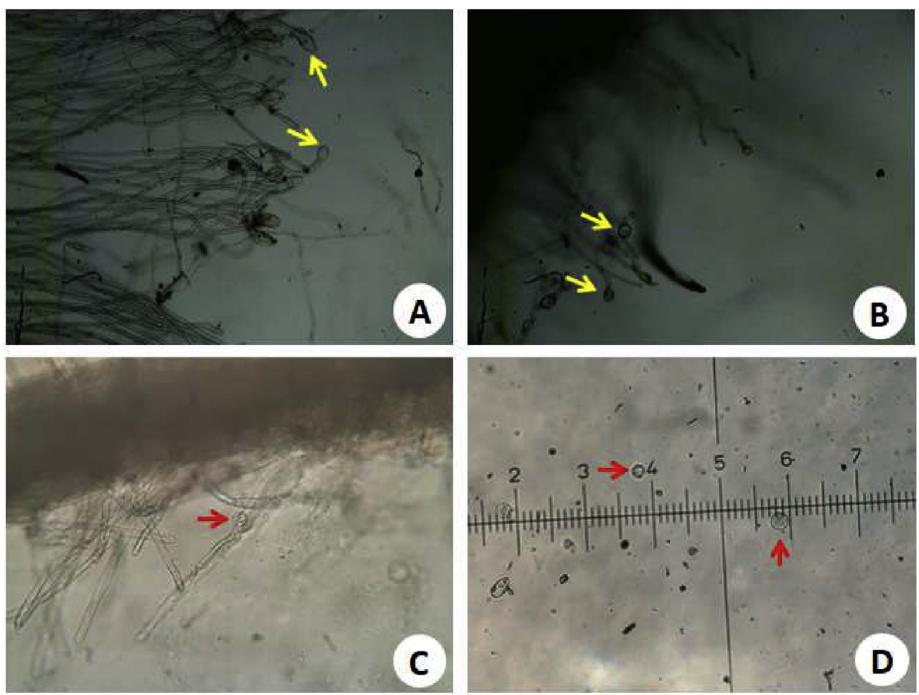

Fig. 2. Stages of isolating transgenic protoplasts from induced hairy roots on the leaves of B. oleracea. A. Swelling of the tops of the root hairs (Yellow arrows) after 3 hours of treatment with the enzyme solution. B. Some protoplasts (Yellow arrows) were released from the tops of root hairs after 8 hours (X10). C. Protoplast isolated from the top of root hairs after enzymatic digestion (Red arrows). D. Viability-retaining transgenic protoplast isolated from hairy roots (Red arrows).

Table 5. The yield, viability, and nucleation percent of transgenic protoplasts isolated from different hairy root cultures induced from the leaves of B. oleracea. Dissimilar letters indicate the presence of significant differences, while the similar letters indicate no significant differences.

| Samples | Yield ($\times 10^4$ cell ml$^{-1}$) | Viability (%) | Nucleation (%) |
| --- | --- | --- | --- |
| 1 | 1.0 c | 64 c | 32 b |
| 2 | 2.0 a | 72 a | 48 a |
| 3 | 1.7 b | 70 b | 47 a |

## 5. Conclusion

One of the most important findings to appear from this study is that the protoplasts are an excellent experimental system for the circulation and generation of genetically modified plants. The enzyme solution of 1.5 percent Cellulase YC and 0.1 percent Pectolyase Y23 at 13 percent mannitol gave a high protoplast yield with $2 \times 10^4$ cells ml$^{-1}$, also the protoplast viability was 72 percent. A PCR assay has confirmed the presence of *rol B* and *rol C* genes in the hairy root tissues of *B. oleracea* L. induced by the *A. rhizogenes* strain (ATCC13332). The manuscript reveals that the *A. rhizogenes* strain is a remarkable vector for incorporating T-DNA genes into cauliflower plants. The protoplast produced from the tissues of hairy roots is considered genetically transformed because it is produced from genetically transformed cells. Perhaps these cells that have lost their cell walls may be invested in the production of genetically transformed plants or the production of genetically transformed callus from protoplasts isolated from plant cells. Also, the adoption of somatic hybridization through fusion between protoplasts isolated from these roots may provide a new path in obtaining plants with desirable features.

In general, fusion is one of the critical paths of genetic transformation in plants. In the future, protoplast fusion can be used to combine this genetically transformed protoplast with another protoplast to obtain genetically transformed plants, especially from those plant species that are challenged to transform by using the standard methods.

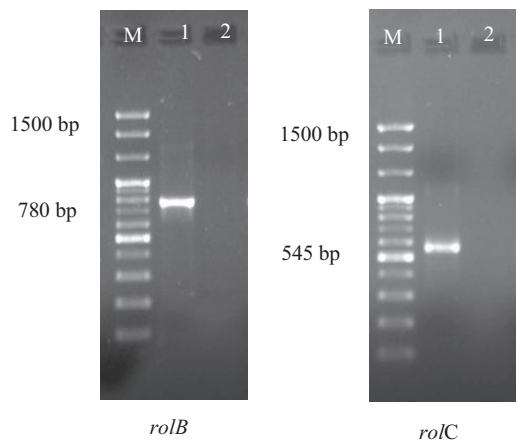

Fig. 3. Agarose gel electrophoresis of amplified DNA samples extracted from transgenic hairy roots and normal roots of Brassica oleracea. Agarose gel image showing amplicon of rol B and rol C genes. Lane (1) 100 bp DNA ladder, lane (2) Amplifier DNA isolated from transgenic hairy roots, lane (3) Amplifier DNA isolated from normal hairy roots.

## Acknowledgement

The authors would like to thank the University of Mosul/College of Education for Pure Sciences for their facilities, which have helped enhance the quality of this work.